\documentclass[aps,groupedaddress, prd,twocolumn]{revtex4}
\usepackage{graphicx,epsf,amssymb,amsmath,latexsym,epsfig}
\newcommand\pa{\partial}

\newcommand\be{\begin{equation}}
\newcommand\ee{\end{equation}}

\begin{document}

\title{Instability of Axions and Photons In The Presence of Cosmic Strings}

\author{Eduardo I. Guendelman}
\email{guendel@bgu.ac.il}
\author{Idan Shilon}
\email{silon@bgu.ac.il}

\affiliation{Physics Department, Ben-Gurion University of the Negev, Beer-Sheva 84105, Israel}

\begin{abstract}
We report that axions and photons exhibit instability in the presence of cosmic strings that are carrying magnetic flux in their core. The strength of the instability is determined by the symmetry breaking scale of the cosmic string theory. This result would be evident in gamma ray bursts and axions emanating from the cosmic string. These effects will eventually lead to evaporation of the cosmic string.
\end{abstract}

\maketitle

\setcounter{equation}{0}

The possible existence of a light pseudo scalar particle is a very interesting possibility. For example,
the axion \cite{Peccei} - \cite{Wilczek},  which was introduced in order to solve the strong CP problem, has since then also been postulated as a candidate for the dark matter. A great number of ideas and experiments for the search this of particle have been proposed \cite{Goldman}, \cite{Review}.

Related to that, in a series of recent publications \cite{duality} one of us showed that an axion-photon system displays a continuous axion - photon duality symmetry when an external magnetic field is present and when the axion mass is neglected. This allows one to analyze the behavior of axions and photons in external magnetic fields in terms of an axion-photon complex particle. For example, the deflection of light from magnestars has been recently studied using these techniques \cite{doron}.

Together with this, the cosmic string solutions which contain a magnetic flux in their core have been extensively studied \cite{vilenkin}. In particular, we concentrate here on cosmic strings that are generated by breaking of a local $U(1)$ symmetry in the abelian Higgs model. This admits string solutions in the form of the Nielsen and Olesen vortex lines \cite{nils}. 

In this letter we show that the coupling of axion-photon complex particles to the magnetic flux of the cosmic string renders the cosmic string unstable. This results in strong gamma ray bursts away from the cosmic string, which would make the existence of cosmic strings extremely prominent.

To see this, let us write the Lagrangian describing the relevant light pseudoscalar coupling to the photon,

\begin{equation}
\label{axion photon ac }
\begin{array}{c}
	\mathcal{L} =  
	 -\frac{1}{4}F^{\mu\nu}F_{\mu\nu} + \frac{1}{2}\partial_{\mu}\phi \partial^{\mu}\phi - 
	\frac{1}{2}m_{a}^{2}\phi^{2} -  \vspace{4pt}\\
	- \frac{g}{8} \phi \epsilon^{\mu \nu \alpha \beta}F_{\mu \nu} F_{\alpha \beta} ~,
\end{array}
\end{equation}

and, following Ref. \cite{idan} (and references therein), specialize to the case where we consider an electromagnetic field with propagation along the $x$ and $y$ directions and a strong magnetic field pointing in the $z$-direction to be present. The magnetic field may have an arbitrary space dependence in $x$ and $y$, but it is assumed to be time independent.

For the small perturbations, we consider only small quadratic terms in the Lagrangian for the axion and the electromagnetic fields, but now considering a static magnetic field pointing in the $z$ direction having an arbitrary $x$ and $y$ dependence and specializing to $x$ and $y$ dependent electromagnetic field perturbations and axion fields. This means that the interaction between the background field and the axion and photon fields reduces to
 
\begin{equation}
\label{axion photon int }
	\mathcal{L}_I =  - 
	 \beta \phi E_z ~,
\end{equation}

where $\beta = gB(x,y) $. Choosing the temporal gauge for the photon excitations and considering only the $z$-polarization for the electromagnetic waves (since only this polarization couples to the axion) we get the following 2+1 dimensional effective Lagrangian

\begin{equation}
\label{2 action}
	\mathcal{L}_{2} =  
	 \frac{1}{2}\partial_{\mu}A \pa^{\mu}A+ \frac{1}{2}\partial_{\mu}\phi \pa^{\mu}\phi - 
	\frac{1}{2}m_{a}^{2}\phi^{2} + \beta \phi \pa_{t} A ~,
\end{equation}

where $A$ is the $z$-polarization of the photon, so that $E_z = -\partial_{t}A$.  

Without assuming any particular $x$ and $y$-dependence for $\beta$, but still insisting that 
it will be static, we see that in the $m_{a}=0$ case (the validity of this assumption will be discussed at the end of this report), we discover a continuous axion photon duality symmetry. This is due to a rotational $O(2)$ symmetry in the axion-photon field space, allowed by the axion and photon kinetic terms, and also since the interaction term, after dropping a total time derivative, can also be expressed in an $O(2)$ symmetric way as follows:

\begin{equation}
\label{axion photon int2}
	\mathcal{L}_I =
	\frac{1}{2}\beta(\phi \pa_{t} A - A \pa_{t}\phi) ~.
\end{equation}

Defining now the axion-photon complex field, $\Psi$, as

\begin{equation}
\label{axion photon complex}
	\Psi = \frac{1}{\sqrt{2}}(\phi + iA) 
\end{equation}

and plugging this into the Lagrangian results in

\begin{equation}
\label{ }
	\mathcal{L} = \partial_{\mu}\Psi^{*}\partial^{\mu}\Psi -\frac{i}{2}\beta(\Psi^{*}\partial_{t}\Psi - \Psi\partial_{t}	\Psi^{*}) ~,
\end{equation}

where $\Psi^{*}$ is the charge conjugation of $\Psi$. From this we obtain the equation of motion for $\Psi$

\begin{equation}
\label{equation7}
	\partial_{\mu}\partial^{\mu}\Psi + i\beta\partial_{t}\Psi = 0 ~.
\end{equation}

Writing separately the time and space dependence of axion-photon field as $\Psi = \mbox{e}^{-i\omega t}\psi(\vec{x})$ and considering the magnetic field of an infinitely thin cosmic string lying along the $z$ axis, reduces Eq. (\ref{equation7}) to

\begin{equation}
\label{eom}
	[-\vec{\nabla}^{2} + gB_{0}\omega\delta(x)\delta(y)] \psi(\vec{x}) = \omega^{2} \psi(\vec{x}) ~,
\end{equation}

where $B_{0}$ is magnetic flux of the cosmic string. Transforming to Fourier space,

\begin{equation}
\label{ }
	\vec{k}^{2}\phi(\vec{k}) + gB_{0}\omega \psi(0) = \omega^{2}\phi(\vec{k}) ~,
\end{equation}

yields the solution

\begin{equation}
\label{ }
	\phi(\vec{k}) = - \frac{gB_{0}\omega\psi(0)}{\vec{k}^{2} - \omega^{2}} ~.
\end{equation}

Following Thorn \cite{thorn}, who solved a similar problem of a non relativistic Schr\"odinger equation with a two dimensional delta function, we integrate the latter over $\vec{k}$ 

\begin{equation}
\label{ }
	\int \frac{d^{2}k}{(2\pi)^{2}}\phi(\vec{k}) = \psi(0) = -\int \frac{gB_{0}\omega\psi(0)}{\vec{k}^{2} - \omega^{2}}			\frac{d^{2}k}{(2\pi)^{2}}~,
\end{equation}

to obtain an eigenvalue condition on $\omega^{2}$

\begin{equation}
\label{ }
	1 = -\frac{gB_{0}\omega}{4\pi^{2}} \int \frac{d^{2}k}{\vec{k}^{2} - \omega^{2}} ~.
\end{equation}

In order to stop this integral from diverging we introduce a cutoff at $|\vec{k}| = \Lambda$. Hence,

\begin{equation}
\label{ome}
	1 = -\frac{gB_{0}\omega}{4\pi}\ln\left(1 - \frac{\Lambda^{2}}{\omega^{2}}\right) \approx -\frac{gB_{0}\omega}{4\pi}\ln	\left(- \frac{\Lambda^{2}}{\omega^{2}}\right) ~,
\end{equation}

where in the last step we assume $\Lambda \gg |\omega|$. By manipulating the latter to the form

\begin{equation}
\label{ome1}
	\frac{2\pi}{B_{0}g\omega}\exp\left({\frac{2\pi}{B_{0}g\omega}}\right) = \frac{2\pi i}{B_{0}g\Lambda} ~,
\end{equation}

we find that $\omega$ is described by Lambert's W function \cite{knuth} 

\begin{equation}
\label{ }
	\omega = \frac{2\pi}{B_{0}g ~ W\left(\frac{2\pi i}{B_{0}g\Lambda}\right)} ~,
\end{equation}

where $W(z)$ satisfies $z = W(z) \mbox{e}^{W(z)}$. Since the $W$ function has an imaginary argument, $\omega$ \textit{must} be a complex function. Therefore, the axion-photon complex particles will excess a (time) instability which will result in axion and photon bursts away from the cosmic string, thus making the string unstable.

Turning now to estimate the strength of the instability, we denote the term $B_{0}g/2\pi$ by $\eta$ and evaluate the magnitudes of $\eta$ and $\Lambda$.

Recent results from the CAST collaboration, that searches for solar produced axions, has set an upper limit on the magnitude of the axion-photon coupling constant of $g < 8.8 \times 10^{-11} ~\mbox{GeV}^{-1}$ for an axion mass of $m_{a} \lesssim 0.02 ~\mbox{eV}$ \cite{g}. Along with this, in Planck units the magnetic flux of a cosmic string is given by $B_{0} = 2\pi n/\sqrt{\alpha}$, where $n$ is an integer, so called the string winding number. Therefore, $\eta = gn/\sqrt{\alpha} \lesssim n \times 10^{-9} ~\mbox{GeV}^{-1}$.

To evaluate the order of the cutoff, $\Lambda$, we understand from dimensional analysis that it is inversely proportional to the only length scale of the system, which is the cosmic string radius. Studying the structure of vortex lines, Nielsen and Olesen \cite{nils} showed that the string width is inversely proportional to the symmetry breaking scale of the theory. Thus, for GUT scale strings, we take $\Lambda \sim 10^{15}~ \mbox{GeV}$.

\begin{figure}[htp]
	\includegraphics[scale=0.45]{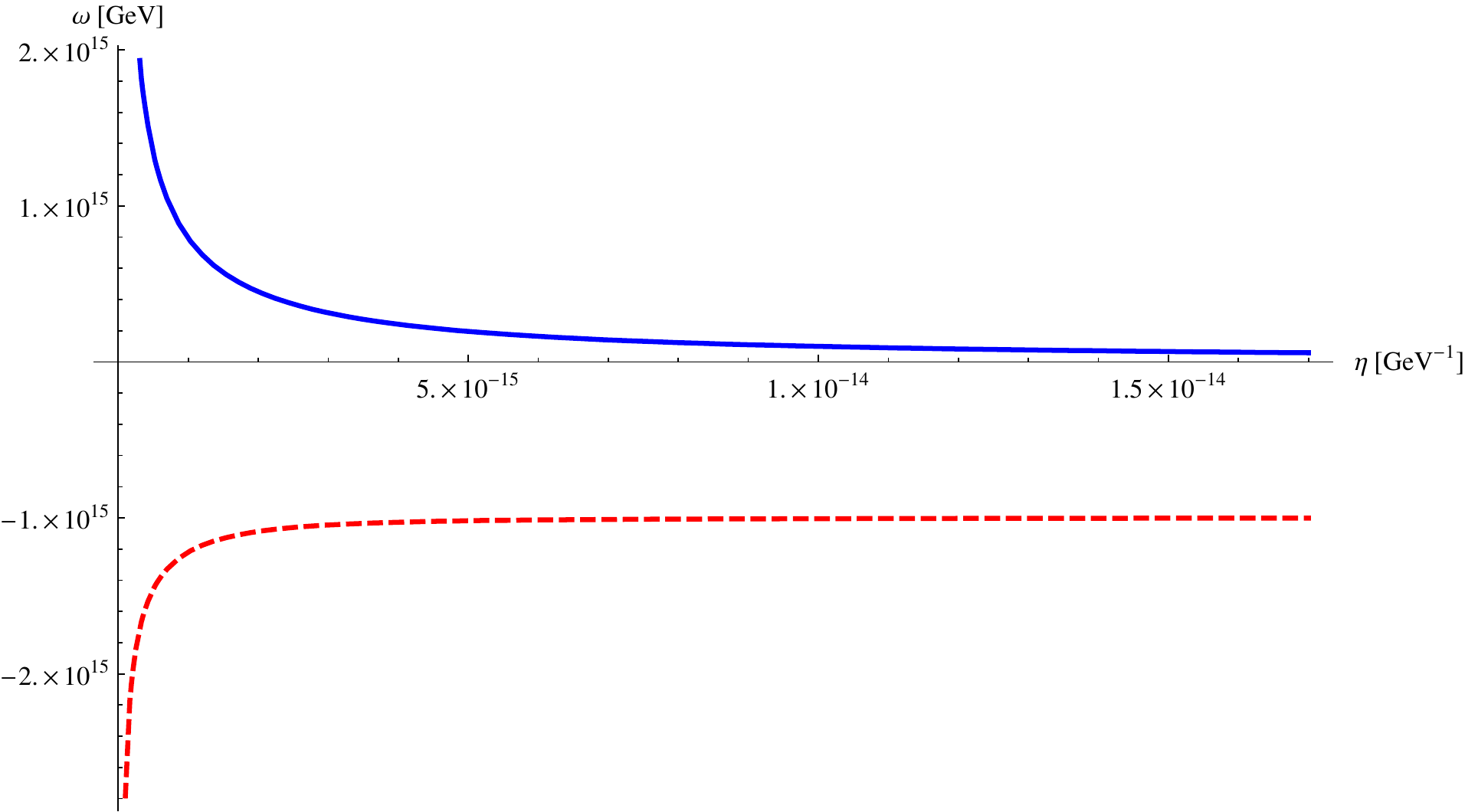}
	\caption{The real and imaginary parts of $\omega$, for $\Lambda = 10^{15} ~\mbox{GeV}$. The real part 	is described by the solid line, while the dashed line represents the imaginary part of $\omega$.  One notes that $	\mbox{Im}[\omega]$ goes to $-\Lambda$ already at the scale $\eta \sim 1/\Lambda$, since $\mbox{Im}[1/W(1.3 i)] \approx 	-1$. On the other hand, it would require a very large value of $\eta$ in order to make the real part of $\omega$ 		negligible (which could be misleading from the graph) since only for $\eta \sim 10 ~\mbox{GeV}^{-1}$, with $		\Lambda \sim 10^{15} ~\mbox{GeV}$, do we get $\mbox{Re}[\omega] \sim 10^{-3} ~\mbox{GeV}		$. For the largest allowed value of $\eta$ (with $n=1$), $\mbox{Re}[\omega] \approx 8.65 \times 10^{7} ~\mbox{GeV}$.}
	 \label{fig:figure1}
\end{figure}

The real and imaginary parts of $\omega$ as functions of $\eta$ are depicted, for $\Lambda = 10^{15}~ \mbox{GeV}$, in Fig. \ref{fig:figure1}.

Taking the limit $\eta \rightarrow \infty$, the asymptotic values of the real and imaginary parts of $\omega$ are given by

\begin{equation}
\label{ }
	\lim_{\eta \rightarrow \infty} \mbox{Re}[\omega] \rightarrow 0 ~,~~ \lim_{\eta \rightarrow \infty} \mbox{Im}[\omega] 	\rightarrow -\Lambda ~.
\end{equation}

Hence, the imaginary part of $\omega$ is bounded from above by $-\Lambda$, rendering the instability extremely strong, for a cutoff of the GUT scale. In particular, this analysis shows that the instability strength is independent of the axion-photon coupling constant magnitude, but is rather determined by the symmetry breaking scale of the cosmic string theory.  

The effective energy of the axion-photon particles is given by $\omega^{2}$. From the properties of $\omega$, we see that the particles effective energy will be complex as well, with a very large and negative real part  (in the order $-1 \times 10^{15} ~\mbox{GeV}^{2}$) and a negligible imaginary part (in the order $10^{-8} ~\mbox{GeV}^{2}$). Hence, as long as the axion-photon particles are localized around the string they will be in a bound state. However, the perturbation will exponentially grow in time eventually causing the cosmic string to evaporate.

The results we have obtained so far are valid for a particle moving on a plane perpendicular to the string axis. We now claim that the same conclusions will be obtained for the more general case of a wave function $\psi(\vec{x})$ with a general momentum as well. To address the general situation, we write Eq. (\ref{ome}) in an invariant form with respect to boosts in the $z$ direction. The electromagnetic field is unchanged by this transformation since the boost is pointing along the direction of the magnetic field. The quantity that will transform under the boost is $\omega$. However, there is an invariant quantity in the form of  $\omega^{2} - k_{z}^{2}$. 

Thus, Eq. (\ref{ome}) in an arbitrary frame becomes

\begin{equation}
\label{ }
	1 = -\frac{gB_{0}\sqrt{\omega^{2} - k_{z}^{2}}}{4\pi}\ln\left(- \frac{\Lambda^{2}}{\omega^{2} - k_{z}^{2}}\right)
\end{equation}

and we find $\omega$ to be

\begin{equation}
\label{ }
	\omega = \left(\frac{4\pi^{2}}{(B_{0}g)^{2}W^{2}\left(\frac{2\pi i}{B_{0}g\Lambda}\right)} + k_{z}^{2}\right)^{1/2} ~.
\end{equation}

Therefore, $\omega$ is complex for any $k_{z}$ and our previous results appear in all modes. 

Lastly, we turn to discuss the validity of the $m_{a} = 0$ approximation we made at the beginning of this letter and show that our conclusions are valid for massive axions as well. In order to verify the massless axion approximation we compare the Compton wavelength of the axion, $\sim 1/m_{a}$, with the localization length of the axion-photon particles wave function. The axion mass is known to be restricted to the region $3 \times 10^{-3} ~\mbox{eV} > m_{a} \gtrsim 10^{-6} \mbox{eV}$ \cite{sik}. Along with this, as one can see from the explicit solution, the wave funcion will be localized around the cosmic string with a size $|1/\sqrt{-\omega^{2}}|$, which is much smaller than the compton wavelength of the axion and therefore making the axion mass irrelevant to our problem. 

In conclusion, we have shown that axionic and electromagnetic excitations will be extremely unstable in the presence of a magnetic flux carrying cosmic string. The axions and the photons will be trapped in a bound state as long as they are localized in the vicinity of the string, but as the perturbation becomes significant extremely rapidly, strong axion and gamma ray bursts will quickly emanate from the string, taking their energy from it and thus bringing its existence to an early end. We also note that our conclusions are true for any value of the axion-photon coupling constant and are determined solely by the symmetry breaking scale of the cosmic string theory.

The same conclusions are of course valid as well for axion like particles, which have no lower bound on their mass and are usually considered very light \cite{ron}, thus making the $m=0$ assumption we made even more valid.


\vskip.3in

\centerline{{\bf Acknowledgment}} 

We would like to thank Doron Chelouche for helpful discussions on the subject.


\end{document}